\documentclass[aoas,preprint]{imsart}

\RequirePackage[OT1]{fontenc}
\RequirePackage{amsthm,amsmath}
\RequirePackage[colorlinks,citecolor=blue,urlcolor=blue]{hyperref}

\RequirePackage{amsfonts,amssymb}
\RequirePackage[authoryear]{natbib}

\arxiv{arXiv:0000.0000}

\startlocaldefs
\numberwithin{equation}{section}
\theoremstyle{plain}

\endlocaldefs

\usepackage[normalem]{ulem}
\usepackage[figuresright]{rotating}
\usepackage{xcolor}
\usepackage{tikz}
\usetikzlibrary{arrows,shapes,backgrounds}
\usepackage{graphicx}
\usepackage{bm, bbold}
\usepackage{amsmath}
\usepackage{mathtools}
\usepackage{hyperref}

\DeclareMathAlphabet{\pazocal}{OMS}{zplm}{m}{n}

\newcommand{\UN}[1]{\text{N}(#1)}

\newcommand{\matN}[1]{\mathcal{M}\mathcal{N}(#1)}
\newcommand{\VEC}[1]{\text{vec}(#1)}

\def\bSig\mathbf{\Sigma}

\newcommand*\diff{\mathop{}\!\mathrm{d}}

\begin{document}

\begin{frontmatter}
\title{Region-Referenced Spectral Power Dynamics of EEG Signals: A Hierarchical Modeling Approach\thanksref{T1}}
\runtitle{Hierarchical Spectral Power Dynamics}
\thankstext{T1}{This work was supported by the grant R01 MH122428-01 (DS, DT) from the National Institute of Mental Health.}

\begin{aug}
\author{\fnms{Qian} \snm{Li}\thanksref{m1}\ead[label=e1]{qianl@ucla.edu}},
\author{\fnms{John} \snm{Shamshoian}\thanksref{m1}},
\author{\fnms{Damla} \snm{\c{S}ent\"{u}rk}\thanksref{m1}},
\author{\fnms{Catherine} \snm{Sugar}\thanksref{m1}},
\author{\fnms{Shafali} \snm{Jeste}\thanksref{m1}},
\author{\fnms{Charlotte} \snm{DiStefano}\thanksref{m1}},
\and
\author{\fnms{Donatello} \snm{Telesca}\thanksref{m1}\ead[label=e2]{dtelesca@ucla.edu}}

\runauthor{Q. Li et al.}

\affiliation{University of California, Los Angeles\thanksmark{m1} }

\address{Donatello Telesca\\
Department of Biostatistics\\
UCLA Fielding School of Public Health\\
Los Angeles, CA\\
\printead{e2}}

\end{aug}

\begin{abstract}
Functional brain imaging through electroencephalography (EEG) relies upon the analysis and interpretation of high-dimensional, spatially organized time series.  We propose to represent time-localized frequency domain characterizations    of EEG data as region-referenced functional data. This representation is coupled with a hierarchical regression modeling approach to multivariate functional observations. Within this familiar setting, we discuss how several prior models relate to structural assumptions about multivariate covariance operators.  An overarching modeling framework, based on  infinite factorial decompositions, is finally proposed to balance flexibility and efficiency in estimation.  The motivating application stems from a study of implicit auditory learning, in which typically developing (TD) children, and children with autism spectrum disorder (ASD) were exposed to a continuous speech stream.  Using the proposed model, we examine differential band power dynamics as  brain function is interrogated throughout the duration of a computer-controlled experiment. Our work offers a novel look at previous findings in psychiatry, and provides further insights into the understanding of ASD. Our approach to inference is fully Bayesian and implemented in a highly optimized Rcpp package. 
\end{abstract}


\end{frontmatter}

%
%
\section{Introduction}
\label{s: intro}
The human brain and its functional relation to  biobehavioral processes like motor coordination, memory formation and perception, as well as pathological conditions like Parkinson's disease, epilepsy, and autism have been a subject of intense scientific scrutiny \cite{broyd2009default}, \cite{uhlhaas2010abnormal}, \cite{murias2007resting}. An important and highly prevalent imaging paradigm aims to study macroscopic neural oscillations projected onto the scalp in the form of electrophysiological signals, and record them by means of electroencephalography (EEG). 

Currently, a typical multi-channel EEG imaging study is carried out with a geodesic EEG-net, composed of up to 256 electrodes. After precise placement on the scalp, electrodes collect electrophysiological signals at high time-resolution, through event-related potentials (ERP) or event-related oscillations (ERO).  In this setting, endogenous or exogenous events can result in frequency-specific changes to ongoing EEG oscillations. The spectral features of the resulting time-series are often used to quantify such changes. Specifically, in a time-frequency analysis, the spatiotemporal dynamics of frequency-specific power are examined as they relate to sensory, motor and/or cognitive processes \cite{Scheffler:2018}  \cite{gou2011resting} \cite{mills1993language}.  

Our work is motivated by a study of language acquisition in young children with autism spectrum disorder (ASD), a developmental condition that affects an individual's communication and social interactions \cite{lord2000autism}. It is thought that typically developing (TD) infants, as young as 6 months old, start to parse continuous streams of speech to actively discern word patterns \cite{kuhl2004early}.  Infants diagnosed with ASD, tend to feature late development of linguistic skills  \cite{eigsti2011language}. Because, neither verbal instructions nor behavioral evaluations can be performed on toddlers, EEG platforms have been recognized as an effective and non-invasive functional brain imaging tool.

\begin{figure}[h!]
	\includegraphics[width = 1.0\linewidth]{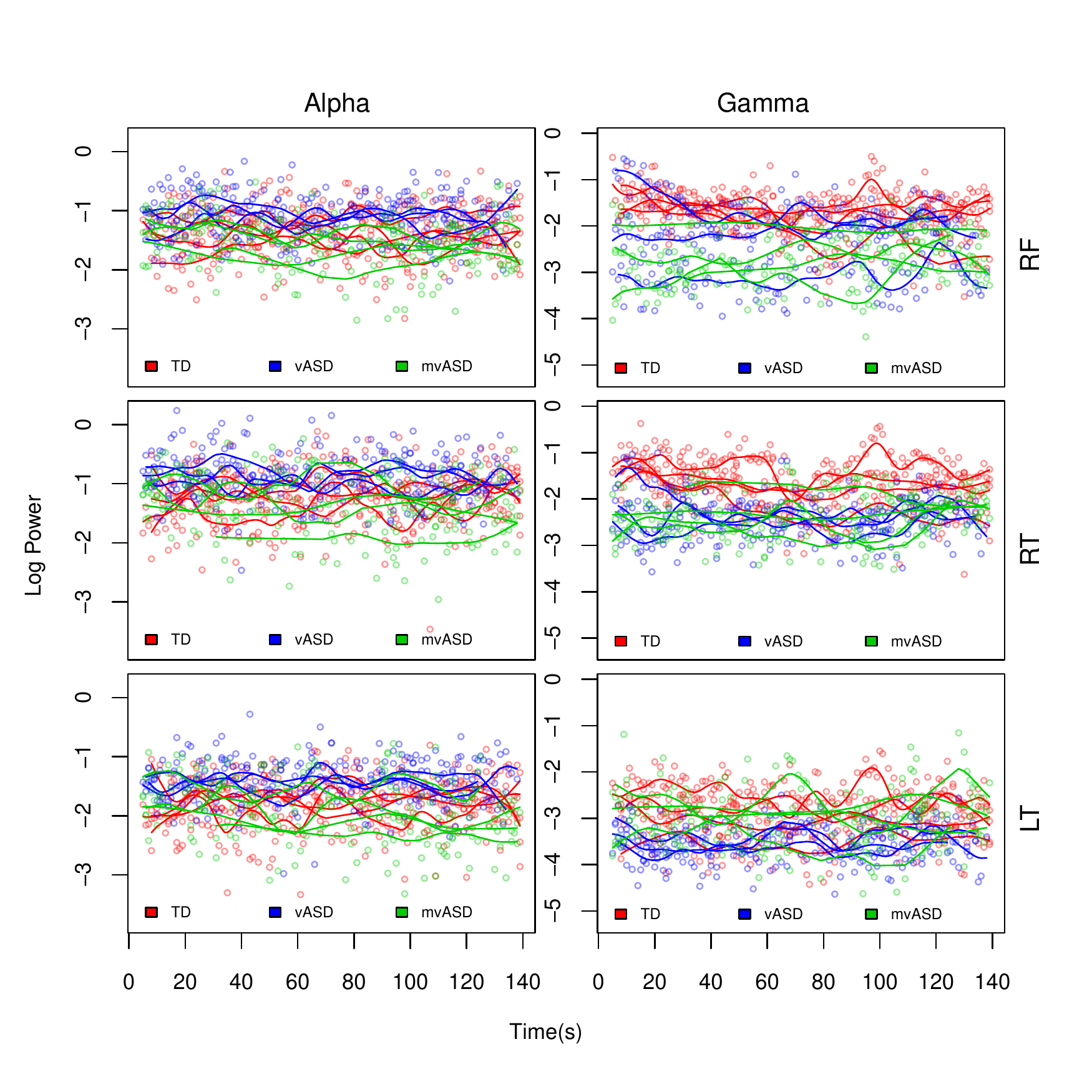}\\
\caption[Data]{\small {\bf Power Dynamics in Language Acquisition.} Log power dynamics both raw and smoothed, in three of eleven anatomical regions: (LT) left temporal, (RT) right temporal, and (RF) right frontal.   For each region, we plot the principal component power for two frequency bands (Alpha, Gamma) during time on trial. We select three subjects from each of the three diagnostic groups: (TD) typically developing, (vASD) verbal ASD, and (mvASD) minimally verbal ASD.}
\label{fig:data}
\end{figure}

In an idealized setting, for each study unit we aim to characterize variability in the relative log-power of a specific frequency band, recorded at a scalp location $s$, at time $t$. This aim is, however, complexed by several estimation- and data-related challenges. Raw EEG recordings, in fact,  suffer from exogenous contamination, poor spatial resolution, as well as non-stationarity. To mitigate these issues, several pre-processing and regularization strategies have been proposed in the literature  \cite{Scheffler:2018}. In this article, we build on our previous work and  couple the characterization of EEG spectral dynamics as region-referenced functional data with a Bayesian model for multivariate functional observations.  Fig.~\ref{fig:data} illustrates the fundamental nature of the data structures considered in our motivating case study, by depicting specific band power trajectories through time-in-trial and across 3 of 11 brain regions in a study of language acquisition.  Given this data structure, we are interested in region-specific, time-varying differences in band power dynamics between diagnostic groups. Crucially, valid inference depends on the correct specification of large multivariate covariance functions.

Many approaches to functional data analysis rely on functional principal components analysis as a fundamental tool for advanced modeling \cite{Yao:2005} \cite{doi:10.1093/biomet/87.3.587}. Methods for the analysis of multivariate functional data are more sporadic, but tend to follow similar strategies with specific attention paid to data scales  \cite{Chiou:2014} or potential heterogeneity in the functional evaluation domain \cite{Happ:2018}. The Bayesian literature on the subject is more sparse, with notable exceptions to be found in  \cite{Baladandayuthapany:2008}, who consider nested spatially correlated functional data under assumptions of separability, and the more general work of \cite{zhou2010reduced}, who consider similar data structures under more flexible covariance models. 

The use of typical spatial assumptions is, however, problematic in the analysis of region-referenced EEG data. Namely, Euclidean distances between electrodes may not accurately characterize functional dependency of within-subject relative log-power readings.  Therefore, instead of relying on pre-specified covariance structures (e.g. conditionally autoregressive models) to describe within-subject correlation, we propose to estimate them through functional factor analysis.

In particular, this article introduces a probabilistic  framework for multivariate functional data  regression which hinges on two ideas, (1) the constructive definition of Gaussian Processes through basis expansions, and (2) the modeling of covariance operators through functional factor analysis \cite{montagna2012bayesian}. While neither approach is new,  we highlight the application of these concepts to functional brain imaging though EEG. In the process, we discuss important consequences associated with specific implementation choices as they relate to transparent assumptions on the structure of multivariate covariance functions. Our proposal seeks to develop a general framework for automatic shrinkage and regularization through infinite factor models  \cite{doi:10.1093/biomet/asr013}. The resulting methodology is suitable for nonparametric estimation with any projection methods,  which is amenable for use in a broad range of applications. 

The article is organized as follows. In Section \ref{sec:preprocessing}, we  outline a conceptual framework for  region-referenced band powers dynamics. A hierarchical functional model for region-referenced functional data is introduced in Section \ref{sec:model}. We assess the performance of the proposed model on engineered datasets in Section \ref{sec:simulation}, and apply our work to a neurocognitive study on auditory implicit learning in Section \ref{sec:casestudy}.  We conclude with a critical discussion of our work in  Section \ref{sec:discussion}.

\section{Time-dependent Region-referenced Band-Power}
\label{sec:preprocessing}
EEG signals, measured through high resolution geodesic networks of electrodes, allow for highly localized interrogation of cortical functions throughout the duration of a computer-controlled experiment. In large studies, the high-dimensional nature of the resulting time series poses potential modeling and computational challenges.  

Due to spatial proximity, neighboring electrodes collect signals likely to be highly multicollinear. In this context, the analysis of EEG data is often preceded by a dimension reduction exercise, aimed at discarding redundant information and aid interpretation through the definition of region-level summaries.  Specifically, electrodes are partitioned according to anatomical regions on the scalp. Subsequently, region-level summaries are defined by averaging power spectra or by selecting the power spectrum of a single electrode within the region. Alternatively, given that multicollinearity of signals defines similarity in their spectral features, spectral PCA \cite{Brillinger:81} has been proposed to pool information within anatomical regions with minimal loss of information. Applications to EEG data have been recently proposed by  \cite{ombao2006time}, and \cite{Scheffler:2018}. Our approach follows closely the development in \cite{Scheffler:2018}, who perform spectral PCA on non-overlapping EEG segments. Our target summaries however focus on specific frequency bands as opposed to the entire power spectrum.  

We outline and define basic concepts used in our work and defer details to Section 2 of the Supplement \cite{Supplement}. 
Let  $\bm{X}_{ij}(s,t)$ denote the raw EEG recording, conceptualized as a locally stationary, zero-mean $q_j$-dimensional random process observed on subject $i$, $i=1,\ldots,n$, within anatomical region $j$, $j=1,\ldots,p$, for segment $t$, $t=1, \ldots, m$, at a sampling rate $U$ across discretized time $s$, $(s = 0, \pm1,\ldots,\pm U/2)$, assuming a 1s analysis window. For each observation $\bm{X}_{ij}(s,t)$, FFT is performed to obtain Fourier coefficients $\bm{d}_{ij}(\omega, t) = U^{-1/2} \sum_{s} \bm{X}_{ij}(s,t)\exp\{-i2\pi\omega s\}$, at frequency $\omega$. The raw periodogram matrix $\bm{I}_{ij}(\omega, t) = \bm{d}_{ij}(\omega,t)\overline{\bm{d}_{ij}}(\omega, t)^{\prime}$ is computed, where $\overline{\bm{d}_{ij}}(\omega, t)^{\prime}$ is the transpose of the complex conjugate of $\bm{d}_{ij}(\omega,t)$.  Following \cite{ombao2006time}, we define kernel smoothed spectral density matrices $\tilde{\bm{I}}_{ij}(\omega, t)$, by smoothing $\bm{I}_{ij}(\omega, t)$ across $\omega$, using  a Daniell kernel with bandwidths selected through generalized cross validation (GCV).  Crucially, smoothing bandwidth are held fixed within region, resulting in  $\tilde{\bm{I}}_{ij}(\omega, t)$ being Hermitian and non-negative definite. 

A one-dimensional region-level summary may be obtained by defining the \emph{principal power} $\lambda^*_{ij}(\omega, t) := c_{ij}^{-1}(t)\,\max_{|\bm{z}|=1} \bm{z}^\prime \tilde{\bm{I}}_{ij}(\omega, t) \,\bm{z}$, as the normalized leading eigenvalue of $\tilde{\bm{I}}_{ij}(\omega, t)$, with $c_{ij}(t) = \int \lambda_{ij}^*(\omega, t)\, d\omega$.  Within region, the principal power summarizes common frequency-level variation across electrodes along the direction of the leading eigenvector.  In contrast to \cite{Scheffler:2018}, who consider region-referenced longitudinal-functional representations, we adopt a simplified view and focus on specific frequency bands.  Specifically, for a frequency range $(\omega_a, \,\omega_b)$, this article emphasizes modeling the \emph{time-varying principal band power} $\gamma_{ij}(t) := \int_{\omega_a}^{\omega_b} \lambda^*_{ij}(\omega, t)\,d\omega$. 

In adults, typical frequency bands and their spectral boundaries are \emph{delta} ($<$4 Hz), \emph{theta} (4-8 Hz), \emph{alpha} (8-15 Hz), \emph{beta} (15-32 Hz), and \emph{gamma} (32-50 Hz). This view is motivated by traditional approaches in neurocognitive science, which differentiate functionally distinct frequency bands as they are thought to target distinct neurocognitive and biobehavioral processes \cite{Saby:2012}.

\section{Hierarchical Models for Region-referenced Functional Data}
\label{sec:model}

\subsection{Data Projections}
Given a frequency band of interest, let $Y_{ij}(t) = \log \{\gamma_{ij}(t)\}$ be the logarithm of the principal band power for subject $i$, anatomical region $j$, evaluated at time $t$.  In practice, for each subject, only a finite number of observations are collected discretely at $\bm{t}_i = \{t_{i1},t_{i2},\ldots,t_{im_i}\}$.  However, for ease of notation, and without loss of generality, we assume $t \in \mathcal{T} = [a, b]$. Collecting all region-level log band-power measurements into a $p$-dimensional vector $\bm{Y}_i(t) = \{Y_{i1}(t), \ldots, Y_{ip}(t) \}'$, we characterize subject-level observations as possibly contaminated multivariate functional data. Specifically, let $\bm{f}_i(t) = \{f_{i1}(t), \ldots, f_{ip}(t)\}'$ be a set of $p$-dimensional random functions, with each $f_{ij}(t)$, $j=1,\ldots, p$, in $\mathcal{L}_2(\mathcal{T})$, a Hilbert space of square integrable functions with respect to the Lebesque measure on $\mathcal{T}$. We assume,
\begin{equation}
	\bm{Y}_i(t) = \bm{f}_i(t) + \bm{\epsilon}_i(t), \quad \bm{\epsilon}_i(t)\sim N_p(0,\bm{\Sigma_\epsilon}), \label{eqn:yfe}
\end{equation}
with residual covariance $\bm{\Sigma}_\epsilon = \text{diag}(\sigma_{\epsilon 1}^2, \ldots, \sigma_{\epsilon p}^2)$. Given $\bm{f}_i(t)$, at time $t$, $\bm{Y}_i(t)$ is assumed to arise from independent realizations of a heteroscedastic multivariate Gaussian distribution. 

We characterize the random functions $\bm{f}_i(t)$ through finite-dimensional projections onto $q$ cubic $B$-spline bases $\bm{b}(t) = \{b_1(t),\ldots,b_q(t)\}^\prime$. Specifically, let $\bm{\Theta}_i = \{\theta_{ijk}\}\in\mathbb{R}^{p\times q}$ be a matrix of random basis coefficients, we express \ref{eqn:yfe} as follows:
\begin{equation}
    \bm{Y}_i(t) = \bm{\Theta}_i \bm{b}(t)' + \bm{\epsilon}_i(t). \label{eqn:datalvl}
\end{equation}

The representation above is readily available for hierarchical modeling.  Given priors on $\bm{\Theta}_i$ and $\bm{\Sigma}_\epsilon$, standard posterior inference and computation applies to the seemingly challenging problem of modeling multivariate functional data.
Crucially, if the elements of $\bm{\Theta}_i$ are assumed Gaussian, given $\bm{b}(t)$, the simple construction  in \ref{eqn:datalvl} implies that $\bm{f}_i(t)$ follows a $p$-dimensional Gaussian process, with second-order properties fully determined by covariance assumptions on the random basis coefficients. 

In the following sections we develop an overarching structure for prior specification, including possible dependence on covariate information, and discuss modeling strategies and their implications for flexibility and efficiency in estimation.

\subsection{Hierarchical Priors and Dependence on Covariates}
\label{subsec:Prior}
A typical study of functional brain imaging aims to relate subject-level time-stable covariate information $\bm{w}_i = (w_{i1},\ldots, w_{id})'$, 
with a region-referenced dynamic outcome $\bm{Y}_i(t)$. We focus on the conditional expectation $E\{\bm{Y}_i(t) \mid \bm{w}_i\} = E(\bm{\Theta}_i \mid \bm{w}_i)\,\bm{b}(t)^\prime$ as the principal measure of association between covariates and brain function outcomes.  Let $M(\bm{w}_i) \coloneqq E(\bm{\Theta}_i \mid \bm{w}_i)$; a familiar modeling framework for $\bm{\Theta}_i$  relies on the following matrix mixed effects construction,
\begin{equation}
	\bm{\Theta}_i = M(\bm{w}_i) + \bm{Z}_i, \quad \VEC{\bm{Z}_i} \sim N_{pq}(0, \bm{\Sigma}_z); \label{eqn:coeflvl}
\end{equation}
where $\bm{Z}_i\in \mathbb{R}^{p\times q}$ captures subject specific regional and functional variation.  Assuming $\bm{w}_i$ is organized as a regression design vector,  let $\bm{\Psi}_\ell \in \mathbb{R}^{p\times q}$, $\ell=1,\ldots,d$,  be regression coefficients; a matrix linear model defines $M(\bm{w}_i) = \sum_{\ell=1}^d \bm{\Psi}_{\ell}\,w_{i\ell}$. In this setting, the regression functions $\bm{\mu}_\ell(t) \coloneqq \bm{\Psi}_\ell\bm{b}(t)'$, are  $p$-dimensional varying coefficients, to be interpreted in relation to the design structure encoded in $\bm{w}_i$ \cite{Zhu:2012}.

Given a basis projection, the second order behavior indexing functional and region-level dependence for the the multivariate Gaussian process $\bm{f}_i(t)$ is fully determined by the covariance matrix $\bm{\Sigma}_z$. While finite-dimensional, this matrix is likely high-dimensional as its spans both the number of regions $p$, and a potentially large number of basis functions $q$. We discuss three modeling approaches, aimed at regularization through shrinkage and simplifying structural assumptions, namely: (NB)  a Na\"{i}ve-Bayes prior, (SS) a Separable Sandwich prior,  and (NS) a regularized Non-Separable prior. 

\vskip.1in
\noindent \emph{(NB) Na\"{i}ve-Bayes Prior}.  This construction exploits two simplifying assumptions: (1) the random coefficients matrix $\bm{Z}_i$ is assumed to follow a Matrix Gaussian distribution, and (2) covariance along time is parametrized through Bayesian P-splines smoothing penalties  \cite{lang2004bayesian}. Let $\bm{\Omega}_0\in \mathbb{R}^{q\times q}$ be a deterministic penalty matrix, and $\bm{S}\in\mathbb{R}^{p\times p}$ be a covariance matrix indexing dependence between anatomical regions. The \emph{NB} prior  assumes
\begin{equation}
	\bm{Z}_i \sim \matN{0, \bm{S}, \bm{\Omega}_0^{-1}}, \quad\quad \bm{S}^{-1} \sim W(\nu,\bm{S}^{-1}_0/\nu).\label{eq:naive}
\end{equation}
The prior above  structures $\bm{\Sigma}_z = \bm{S}\otimes\bm{\Omega}_0^{-1}$, implying separability of covariation between regions and time points.  The matrix $\bm{S}$ serves both the purpose of modeling dependence between brain regions through its off diagonal elements, and the purpose of establishing region-level adaptive smoothing through its diagonal elements as they multiply $\bm{\Omega_0^{-1}}$.  More details about this matrix and the choice of hyperparameters are discussed in Web Appendix C. While greatly simplified, when compared to a completely unstructured $\bm{\Sigma_z}$, this construction is potentially problematic as it hinges on a relatively rigid parametrization for the time-covariance,  while enforcing only mild Wishart regularization of $\bm{S}$. 

\vskip.1in
\noindent\emph{(SS) Separable Sandwich Prior}. 
A more balanced approach to the prior in \ref{eq:naive} retains the assumption of separability, but implements adaptive regularized estimation of both the time and region covariance. In particular, we extend the approach of \cite{BIOM:BIOM1788}, and propose a two-way Bayesian latent factor model for $\bm{Z}_i$.  Let $\bm{\Upsilon}=\{\upsilon_{jr}\}\in\mathbb{R}^{p\times k_1}$, and $\bm{\Gamma}=\{\gamma_{vs}\}\in\mathbb{R}^{q\times k_2}$ be two loading matrices. Also let $\bm{H}_i = \{h_{rs}\}\in \mathbb{R}^{k_1\times k_2}$, be a random matrix with $h_{rs}\sim_{iid} N(0,1)$, $r=1,\ldots, k_1$, $s=1,\ldots,k_2$, we write:
\begin{equation}
	\bm{Z}_i = \bm{\Upsilon} \bm{H}_i \bm{\Gamma}' + \bm{R}_i, \quad\quad \text{with} \quad \bm{R}_i \sim \matN{\bm{0},\bm{\Sigma}_p, \bm{\Sigma}_q};\label{eqn:SSprior}
\end{equation}
where $\bm{\Sigma}_p$ and $\bm{\Sigma}_q$, both are diagonal with anisotropic components.  This construction, implies
$$\bm{\Sigma_z} =  \bm{\Upsilon}\bm{\Upsilon}' \otimes \bm{\Gamma}\bm{\Gamma}' + \bm{\Sigma}_p \otimes \bm{\Sigma}_q.$$
Typically, as in latent factor models, a truncation $k_1<p$ and $k_2<q$ is selected  to define a low-rank representation of the region and time covariances. Rather than selecting the number of latent factors, we consider the multiplicative shrinkage prior of \cite{doi:10.1093/biomet/asr013}, so that:
\begin{align*}
	\upsilon_{jr} \mid \phi_{jr},\tau_r \sim \UN{0,\phi_{jr}^{-1} \tau_r^{-1}}, &\quad \phi_{jr}\sim \text{Ga}(\nu_1/2, \nu_1/2),
	\quad \tau_r = \prod_{u = 1}^r \delta_u,\\
	\delta_1 \sim \text{Ga}(a_{11},1), &\quad \delta_u \sim\text{Ga}(a_{12},1), \quad \text{when } u>1; \\
	\gamma_{vs} \mid \rho_{vs},\kappa_s \sim \UN{0, \rho_{vs}^{-1}, \kappa_{s}^{-1}}, &\quad \rho_{vs} \sim \text{Ga}(\nu_2/2, \nu_2/2), \quad\kappa_s = \prod_{\ell=1}^s \pi_v, \\
	\pi_1 \sim \text{Ga}(a_{21},1), &\quad\pi_v \sim \text{Ga}(a_{22},1), \quad \text{when }\ell>1.
\end{align*}
When $a_{11}>1, a_{12}>1$, this prior defines stochastically increasing precisions as more columns are added to $\bm{\Upsilon}$ and $\bm{\Gamma}$. Specific details about this shrinkage strategy and choice of hyperparameters are reported in the Supplement (Section 3) \cite{Supplement}. In what follows, we build on latent factor representations and forego the assumption of separability altogether.

\vskip.1in
\noindent \emph{(NS) Non-Separable Prior}.  As noted in \cite{Cressie1999}, the class of separable models (\ref{eq:naive}, \ref{eqn:SSprior}) is severely limited since it cannot capture region-time interaction. The cross-covariance functions between the time series at any region has the same shape, regardless of region location. Separable structures are often chosen for convenience rather than for their ability to fit the data well. These limitations make the separability approach difficult to justify in the setting of functional neuroimaging.  A simple approach, which foregoes this assumption, while still defining a regularized representation of $\bm{\Sigma}_z$   involves dealing with  
$\mbox{vec}(\bm{Z}_i)$ directly. Specifically, let $\bm{\Xi} = \{\xi_{cs}\} \in \mathbb{R}^{pq\times k}$, be a loading matrix, and $\bm{\eta}_i \sim N_k(0, I_k)$, we write
\begin{equation}
	\text{vec}(\bm{Z}_i) =  \bm{\Xi} \,\bm{\eta}_i + \bm{r}_i, \quad\quad \text{with}\quad \bm{r}_i \sim N_{pq}(0,\bm{\Sigma}_r),\label{eqn:NSprior}
\end{equation}
with $\bm{\Sigma}_r = \mbox{diag}(\sigma^2_{r1}, \ldots, \sigma^2_{rpq})$, implying $\bm{\Sigma}_z = \bm{\Xi}\,\bm{\Xi}' + \bm{\Sigma}_r$. This formulation leads to a probabilistic version of the multivariate FPCA of \cite{Chiou:2014}, with normalization handled through explicit assumptions of heteroscedasticity through $\bm{\Sigma}_r$. Echoing the approach used in the sandwich prior, the model is completed with a shrinkage prior on the loading matrix $\bm{\Xi}$, so that
\begin{align*}
	\bm{\xi}_{(s)}^{(c)} \mid \bm{\Omega}_s, \tau_c \sim N\left({\bm{0},\, \bm{\Omega}_s^{-1}\tau_c^{-1}}\right), &\quad \bm{\Omega}_s \sim \text{W}(\nu,\bm{\Omega}_0/\nu), \quad \tau_c = \prod_{l=1}^c \delta_l,\\
	\delta_1 \sim \text{Ga}(a_1,1), &\quad\delta_l \sim \text{Ga}(a_2,1), \quad \text{when } l>1;
\end{align*}
where $\bm{\xi}_{(s)}^{(c)}$ indicates the $s$-th $p$-dimensional block, $s\in\{1,\ldots,q\}$, of the $c$-th column of $\bm{\Xi}$, $c=1,\ldots, k$.  

\vskip.1in
\noindent  It is easy to show that the vectorized model in \ref{eqn:NSprior} encompasses both the \emph{separable sandwich} prior in \ref{eqn:SSprior} and the \emph{na\"{i}ve-Bayes} prior in \ref{eq:naive}. We examine  the finite sample properties of these modeling strategies under several simulation scenarios in the section \ref{sec:simulation}. 

\vskip.1in
\noindent The three priors listed above span a broad range of model flexibility.  We recognize, however, that several other nuanced configurations have been considered in the literature; particularly, when sampling schedules allow for design-driven structural assumptions.  For example, in the context of longitudinal functional data, methodological simplifications to multivariate functional data, include both functional random effects models \cite{Greven:2010} and simplified multivariate functional principal components representations \cite{Park:2015}.   Weakly separable covariance structures have been characterized in \cite{chen2017weak}, bridging the complexity of SS and NS models, by expressing a multi-way covariance as the linear combination of rank-one strongly-separable terms. This idea has found several applications and extensions in the context of longitudinal functional data literature \cite{lynch2018test, scheffler2020hybrid, shamshoian2019bayesian}. While a comprehensive comparison is out of scope for this work, in the context of our case study and through simulations,  we aim to assess how a general NS model can avoid losses in efficiency through adaptive regularization. 

\subsection{Posterior Inference}
Posterior inference is based on Markov Chain Monte Carlo simulations from the target measure.
All prior models discussed in section \ref{subsec:Prior} are amenable to simple Gibbs Sampling implementations.  A highly optimized R package for data manipulation and inference is freely available from \verb'github' at \url{https://github.com/Qian-Li}. 

The MCMC transition schedule implemented in our package is optimized to marginalize out subject by region coefficients whenever possible, therefore limiting the degree of autocorrelation between posterior draws. As is the case with GP-type regression, in applying our method beyond EEG data, some care is needed, as the scalability of simple posterior sampling schemes could be an issue in high-dimensions (large $n$ or $p$).  In these cases, specialized approximation strategies are, however, applicable  as described in \cite{Nishimura:2020}. 

We note that the loading matrix $\bm{\Xi}$ in \ref{eqn:NSprior}, as well as $\bm{\Upsilon}$ and $\bm{\Gamma}$ in  \ref{eqn:SSprior} are not likelihood-identified due to invariance to orthogonal rotations.  Crucially, however: $\bm{\Xi}\bm{\Xi}'$, $\bm{\Upsilon}\bm{\Upsilon}'$ and $\bm{\Gamma}\bm{\Gamma}'$ are all uniquely identified, leading to likelihood-identifiability of $\bm{\Sigma}_z$.  Given posterior samples from the model parameters,  Monte Carlo estimates of all quantities of interest, including simultaneous credible bands are obtained in a relatively straightforward fashion \cite{Baladanda:2005}.   Detailed calculations, including full conditional distributions are reported in  the Supplement \cite{Supplement}.

\section{Experiments on Engineered Data}
\label{sec:simulation}
To evaluate the finite sample performance of the hierarchical model and priors described in Section \ref{sec:model} we carried out an extensive Monte Carlo study considering data generated under several dependence structure (separable, non-separable), dependence strength, sample size, and signal-to-noise ratio scenarios.  The signal-to-noise ratio is defined by $\text{SNR} = \frac{1}{n}\frac{1}{p}\sum_{i=1}^{n}\sum_{j=1}^{p}\int_{\mathcal{T}}f_{ij}(t)^{2}dt/\sigma_{\epsilon}^{2}$. Intuitively, SNR in this context can be intended as an inflation quotient for the error-variance $\sigma^2_\epsilon$. The general goal of these experiments aims to assess how well the mean $M(\bm{w}_i)$ and covariance $\bm{\Sigma}_z$ are recovered under different priors and simulation truths. Further details regarding the data generation scheme are available in  Web Appendix (Section 4).

Let $\hat{\bm{\mu}}_\ell(t)$ and $\widehat{\bm{\Sigma}}_z$ denote posterior mean estimates for their parametric counterparts (Section \ref{subsec:Prior}). To quantify the quality of estimates we consider relative squared errors, defined as follows:
\begin{itemize}
\item \emph{Mean}: average relative squared error across regions and covariates, s.t.  
\[
 \text{RSE}(M) = \frac{1}{dp}\sum_{\ell=1}^d\left[\frac{\int_t \{\hat{\bm{\mu}}_\ell(t) - \bm{\mu}_\ell(t)\}^2 \diff t}{\int_t \bm{\mu}_\ell(t)^2 \diff t}\right]^\prime \bm{1}_p;
\]
\item \emph{Covariance}: Relative squared Frobenius error, st. 
$$\text{RSE}(\Sigma_z) = \frac{\lVert \Sigma_{z} - \widehat{\Sigma}_{z} \rVert_F^2}{\lVert \Sigma_{z}  \rVert_F^2}.$$
\end{itemize}
In relation to these metrics we observe the following behavior. 

\begin{table}[h]
\centering
\caption{{\bf Simulation Study.} Mean and Covariance recovery under Na\"{i}ve-Bayes (NB),  Separable-Sandwich (SS), and Non-Separable (NS) priors. Relative errors are reported under separable and non-separable simulation truths, sample size escalation ($n = 10, 20, 30$) and two signal-to-noise ratio  scenarios ($SNR = 0.2, 1.0$).}
\label{tab:B1}
\begin{tabular}{l rr rr rr}
\hline
\\[-.1in]
          & \multicolumn{2}{c}{$n=10$}      & \multicolumn{2}{c}{$n=20$}      & \multicolumn{2}{c}{$n=50$}      \\
 & Mean    & Cov. & Mean  & Cov. & Mean  & Cov.\\
\hline
\\[-.1in] 
\underline{Separable} & \multicolumn{6}{l}{(SNR $= 0.2$)}        \\ 
NB  & 0.0346& 0.3255 &0.0248 &0.3150 &0.0162 &0.3120\\
SS  &0.0384 &0.3192  &0.0272 &0.3004 &0.0174 &0.2806\\
NS  &0.0386 &0.3380  &0.0272 &0.3117 &0.0174 &0.2900\\[.05in]
& \multicolumn{6}{l}{(SNR $= 1.0$)}        \\
NB  &0.0277 &0.5146 &0.0195 &0.5126 &0.0124 &0.5180\\
SS  &0.0272 &0.1987 &0.0192 &0.1817 &0.0122 &0.1677\\
NS  &0.0272 &0.2595 &0.0192 &0.2294 &0.0122 &0.1915\\[.1in]
\underline{Non-Separable} & \multicolumn{6}{l}{(SNR $= 0.2$)}        \\ 
NB  &0.0382 &0.4217 &0.0276 &0.4189 &0.0183 &0.4226\\
SS  &0.0413 &0.3633 &0.0289 &0.3547 &0.0185 &0.3470\\
NS  &0.0415 &0.3427 &0.0290 &0.3188 &0.0185 &0.2892\\[.05in]
& \multicolumn{6}{l}{(SNR $= 1.0$)}        \\
NB &0.0339 &0.6601 &0.0236 &0.6110 &0.0153 &0.6164\\
SS &0.0340 &0.3412 &0.0235 &0.3357 &0.0152 &0.3321\\
NS &0.0338 &0.2721 &0.0234 &0.2433 &0.0151 &0.2101\\
\hline

\end{tabular}
\end{table}

\vskip.1in
\noindent \emph{Sample size and SNR}. For each of 100 datasets, we simulate $n=10, 20, 50$ individuals in each of 3 diagnostic groups, with functional observations collected over 6 anatomical regions. Each data-set is considered under two signal-to-noise ratio scenarios, SNR$=0.2,1.0$. We assume observations are defined on a common time-grid $\in[0,1]$, with random missing patterns, where between $0$ and $80\%$ of the time-points are discarded to mimic observed data.  Results are summarized in Table \ref{tab:B1}.

All three priors recover the mean structure equally well under both SNR scenarios. Independently of the generating dependence structure, average relative squared errors improve as more subjects are included for analysis. More meaningful differences are observed when we consider recovery of the covariance structure.   When data are generated under separable covariance, \emph{SS} priors exhibit the best performance, with improved accuracy for larger sample sizes. Some loss in efficiency is observed when considering \emph{NS} priors. However, even with only 10 subjects the loss in efficiency is estimated to be only about 30\%, reducing to about  4\% when sample size escalates to $n=50$.  When data are generated from non-separably covarying processes, \emph{NS} priors exhibit the highest efficiency, and are the only model improving meaningfully in accuracy as sample size increases. In all simulation settings, \emph{NB} priors seem to perform poorly in relation to other alternatives.

In summary, our findings agree with well known results in multivariate and longitudinal data analysis \cite{Wakefield}. Consistent recovery of the mean function seems to be relatively insensitive to  the covariance model. Inference and uncertainty quantification about the mean, however, requires correct recovery of the dependence structure encoded in $\bm{\Sigma}_z$. While meaningful gains in efficiency can be achieved if separability assumptions are warranted in applications, the encompassing regularized prior (\emph{NS}) leads to a generally more flexible and efficient  modeling framework, without the need for stringent structural assumptions.

\vskip.1in
\noindent  Extended results, including sensitivity to model specification are reported in the Supplement \cite{Supplement}. While theoretical large sample results are not examined in the manuscript, all simulation results can be reproduced and extended with our R package and supplementary documentation.

\section{EEG and Language Acquisition in TD and ASD Infants}
\label{sec:casestudy}

\subsection{Study Background}
\label{sec:background}
We consider a functional brain imaging study  carried out by our collaborators in the Jeste laboratory at UCLA. The study aims to characterize differential functional features associated with language acquisition in TD and ASD children. EEG data were recorded for 144s using an 128 electrode HydroCel Geodesic Sensor Net for 9 TD, 32 ASD children   ranging between 4 and 12 years of age. The EEG data is divided into non-overlapping segments of 1.024 seconds, producing a maximum of 140 observable segments for each subject at each electrode. Details about data pre-processing are deferred to Web Appendix A. Individual sensors were partitioned between 11 anatomical regions made up of 4-7 electrodes; left and right for the temporal region (LT and RT) and left, right, and middle for the frontal, central, and posterior regions (LF, RF, MF, LC, RC, MC, LP, RP, and MP, respectively).  Region-level power dynamics were estimated as outlined in  Section \ref{sec:preprocessing}. Figure \ref{fig:data} illustrates a sample of smoothed power trajectories within the \emph{alpha}  and  \emph{gamma} frequency bands for both TD and ASD children. ASD children were further classified as verbal ASD (vASD - 14 children) and minimally verbal ASD (mvASD - 19 children) in relation to their verbal Developmental Quotients (vDQ).

The broad scientific goal of this study aims to understand the functional underpinnings associated with language acquisition through the human ability to parse otherwise continuous speech streams into meaningful words. In order to replicate this natural phenomenon, children were exposed to continuous synthetic speech constructed through a collection of 12 phonemes. By defining phoneme triplets  (e.g. \emph{pa-bi-ku}, \emph{da-ro-pi}) as deterministic pseudo-words and exposing children to random continuous permutations of pseudo-words, study subjects were given the opportunity for implicit statistical learning of word segmentation.  During this process, we seek to detect differential neurophysiological response across anatomical regions and oscillation frequencies. 

\begin{figure}
	\includegraphics[width = 1.0\linewidth]{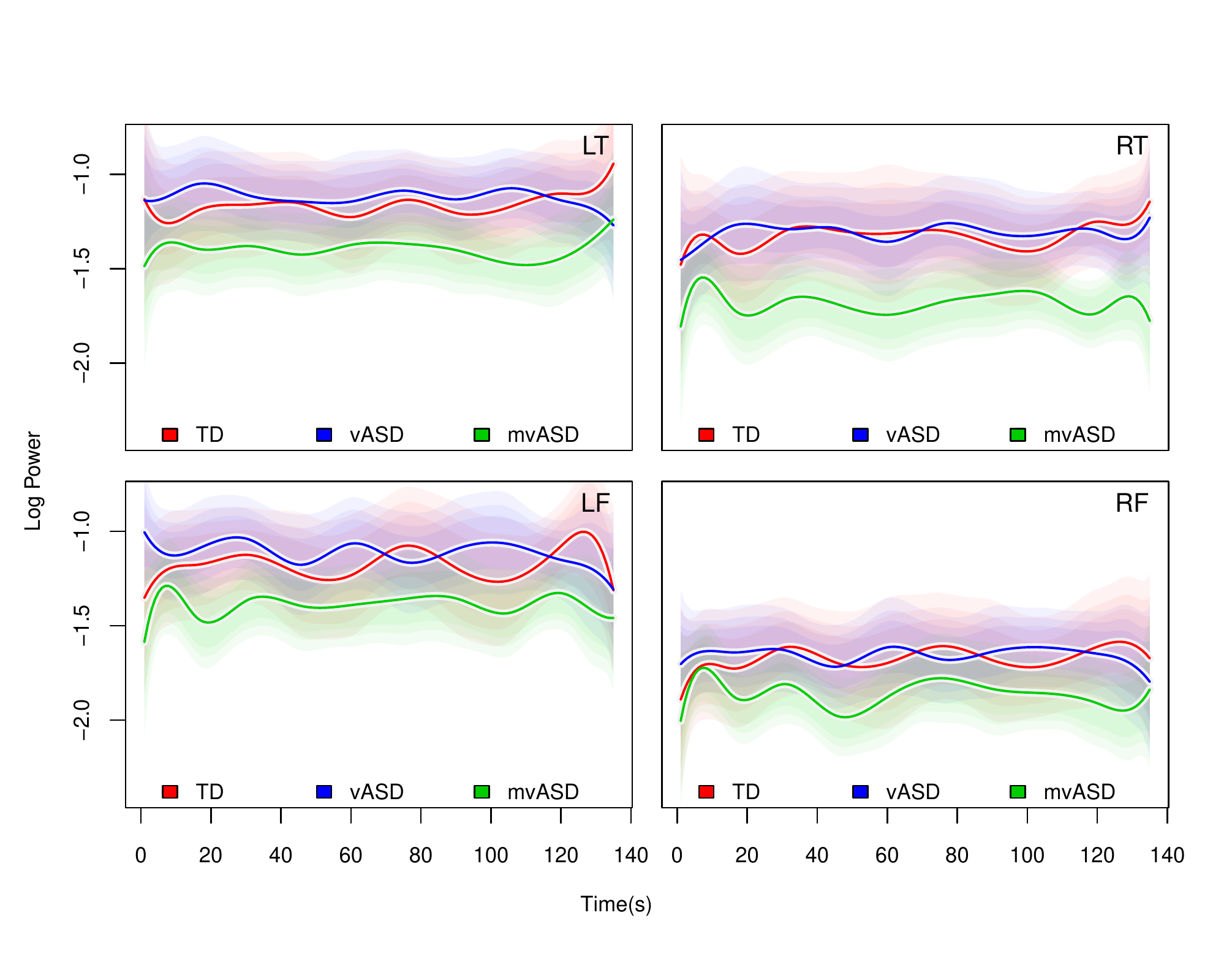}\\
	\caption{\small {\bf Alpha Power Dynamics.} Average log power dynamics stratified by diagnostic group: (TD) typically developing, (vASD) verbal ASD, and (mvASD) minimally verbal ASD.
Four of eleven anatomical regions are represented, namely: (LF) left frontal, (RF) right frontal, (LT) left temporal, and (RT) right temporal. In each region, we display the posterior mean, accompanied by simultaneous credible bands, shaded to include 0.2, 0.6, and 0.8 of all posterior samples.}
\label{fig:alphaGroup}
\end{figure}

\subsection{Group Mean Trajectory Analysis}
\label{sec:group}
Our analysis focuses on time-varying region-referenced differences among diagnostic groups (TD, vASD, mvASD) in relation to two frequency bands, namely  \emph{alpha} (8-15 Hz),  and \emph{gamma} (32-50 Hz). Specifically, alpha waves are thought to play an active role in network coordination and communication \cite{fink2014eeg}, while gamma waves are thought to index conscious perception and  tend to correlate with implicit learning processes \cite{gruber2002effects}.

\begin{figure}[h]
	\includegraphics[width = 1.0\linewidth]{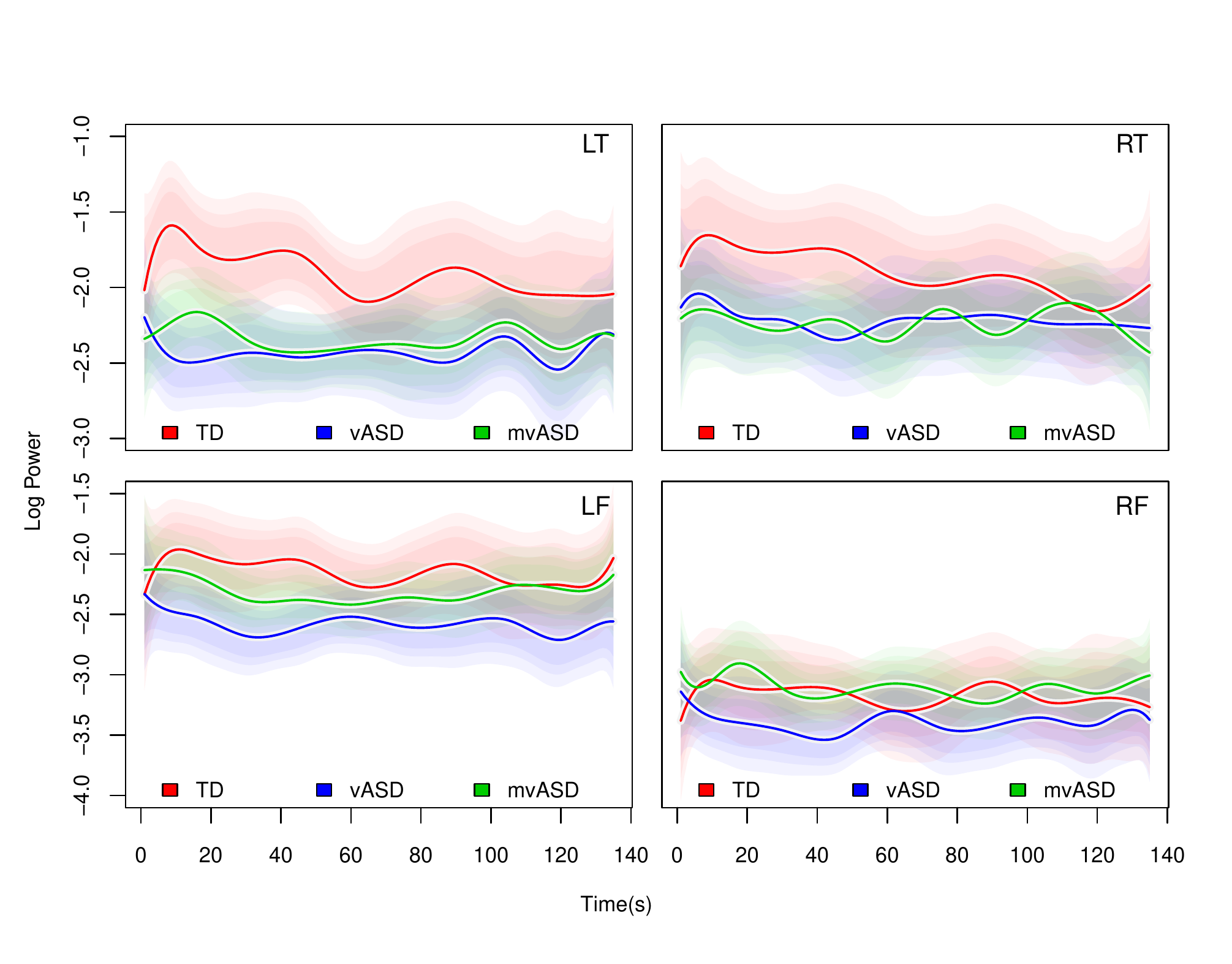}\\
\caption{\small {\bf Gamma Power Dynamics.} Average log power dynamics stratified by diagnostic group: (TD) typically developing, (vASD) verbal ASD, and (mvASD) minimally verbal ASD.
Four of eleven anatomical regions are represented, namely: (LF) left frontal, (RF) right frontal, (LT) left temporal, and (RT) right temporal. In each region, we display the posterior mean, accompanied by simultaneous credible bands, shaded to include 0.2, 0.6, and 0.8 of all posterior samples.}
\label{fig:gammaGroup}
\end{figure}

Our analyses are based on the non-separable model in (\ref{eqn:NSprior}), with 10 latent factors, and data projected on 12 B-splines basis functions. Main findings across 4 of 11 regions are summarized graphically in Fig~\ref{fig:alphaGroup} for alpha waves and in Fig~\ref{fig:gammaGroup} for gamma waves. In both models, the varying coefficients representing the mean structure in \ref{eqn:coeflvl}, take as input a simple factorial covariate indexing diagnostic group membership, and yield group-level means across anatomical regions. For each time-varying mean, we also represent simultaneous credible bands by probability shading, to include a gradient of 0.2, 0.6 and 0.9 posterior coverage.  A brief discussion about sensitivity to the prior model, number of latent factors, and projection methods is deferred to Section \ref{sec:discussion}. 

Starting with a group-by-region analysis of the alpha frequency band in Fig.~\ref{fig:alphaGroup}, we note that for all three groups, across all 11 brain regions, there is no discernible time-varying pattern in the average relative alpha frequency, which is maintained relatively flat across time-in-trial. The TD and vASD groups are essentially indistinguishable, with minor differences likely due to sampling variability. Crucially,  meaningfully lower levels in the average relative alpha frequency are observed for the mvASD group. While we fall short of considering formal notions of statistical significance, we highlight potentially meaningful findings in the RT region, where 90\% simultaneous bands fail to overlap for some of the time-on-trial. More importantly, the lower level in relative alpha frequency is consistent across most brain regions, suggesting a potential role of alpha waves in distinguishing brain functional characteristics in the minimally verbal ASD group.  Finally, within group, for all diagnostic groups, we confirm a well known phenomenon known as left-right alpha asymmetry, which is  significant over time at frontal regions, diminished at temporal regions, and negligible at central and posterior regions. 

Replicating the same analysis for gamma frequencies in Fig.\ref{fig:alphaGroup} we note that, while alpha waves seem to isolate the mvASD group, gamma waves tend to separate TD children from the vASD and mvASD groups. More precisely,  the TD group exhibits significantly higher average gamma band power than vASD and mvASD, especially at the early stages of the experiment. This pattern is most evident at temporal regions, both left and right (LT, RT).  Furthermore, the average gamma band power for TD children exhibits a sharp increase at the beginning of time-on-trial, followed by a decreasing trend through the end of the experiment.  This finding suggests, that differently from the ASD groups, TD children consciously perceive the beginning of the speech stream as a stimulus. Compatible findings by \cite{gruber2002effects} report a decreased response when stimuli recur, which they projected to be linked to a ``neural savings'' mechanism within a cell assembly representing an object,  i.e. a word in our case. As observed for alpha band power, we note significant frontal asymmetry throughout the study and across diagnostic groups. While previous studies, e.g. \cite{rojas2014gamma}, pointed out left-dominant asymmetry as a discriminative feature between TD and ASD cohorts, our findings do not replicate this observation within the context of a language acquisition experiment.

\vskip.1in
\noindent We compared the NB, SS and NS prior models, by computing their expected log pointwise predictive density leave-one-out cross-validation (elpd-loo)\cite{Watanabe:2013, vehtari2017practical, vehtari2015pareto}.  In the context of our case study, let $p\{\boldsymbol{Y}_{i}(t)\mid \boldsymbol{Y}_{-i}(t)\}$ be the leave-one-out predictive density for subject $i$, ($i=1,\ldots,n$).  The elpd-loo is  computed as follows: 
\begin{equation}
    \text{elpd}_{\text{loo}} = \sum_{i=1}^{n}\log p(\boldsymbol{Y}_{i}(t)|\boldsymbol{Y}_{-i}(t))
\end{equation}
Treating gamma band power as the response, NB has $\text{elpd}_{\text{loo}} = -11586$ (SE$=947$), SS with 6 latent factors for the time and spatial domains has $\text{elpd}_{\text{loo}} = -10981$ (SE$=911$), and NS with 10 latent factors has $\text{elpd}_{\text{loo}} = -10951$ (SE$=910$). Treating alpha power as the response, NB has $\text{elpd}_{\text{loo}} = -14698$ (SE$=1227$), SS has $\text{elpd}_{\text{loo}} = -14211$ (SE$=1209$), and NS has $\text{elpd}_{\text{loo}} = -14260$ ($SE=1214$).   For both the gamma and alpha power bands, SS and NS tend have a relatively similar performance, while NB seems to underperform in both settings. Our analyses are therefore based on the NS prior, as it is more general and does not seem to induce any loss in predictive power for this case study.   

\begin{figure}[h]
	\includegraphics[width = 1.0\linewidth]{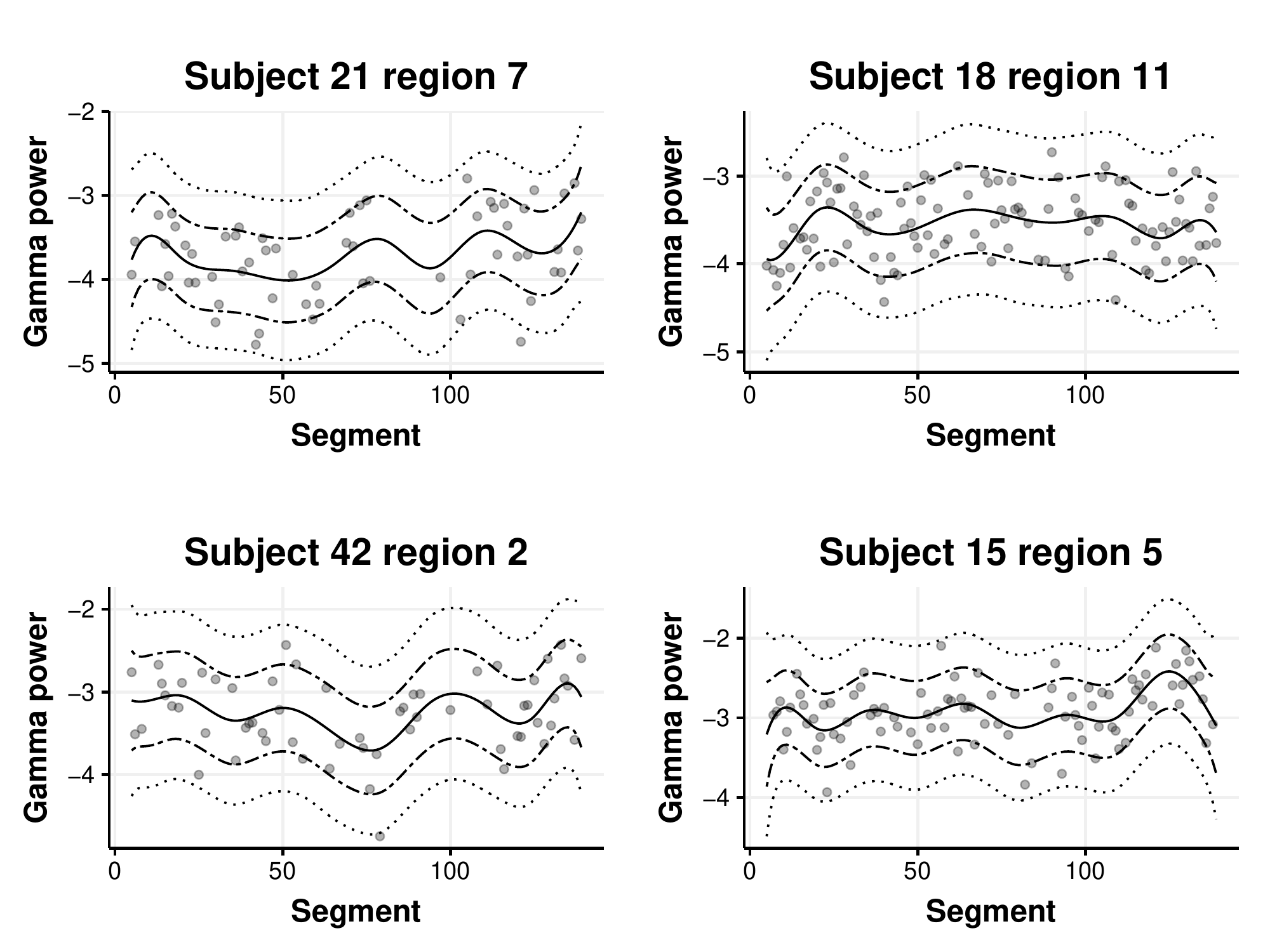}\\
\caption{\small {\bf Gamma power goodness of fit.} Posterior mean trajectory (solid line) for four random subject region responses. We report pointwise (dot-dash lines) and simultaneous (dotted lines) 90\% posterior predictive credible bands.  Top left: subject 21 (mvASD), region 7. Top right:  subject 18 (ASD), region 11. Bottom left: subject 42 (mvASD), region 2. Bottom right:  subject 15 (ASD), region 5.}
\label{fig:fit}
\end{figure}

To assess goodness of fit, we compute 90\% pointwise and simultaneous (grouping by subject and region) posterior predictive credible bands treating gamma power as the response \cite{Gelman:1996}.  Simultaneous bands are obtained as in \cite{Crainiceanu:2007}. Figure \ref{fig:fit} displays four random subject-region gamma power dynamics accompanied with maximum a posteriori fit and 90\% pointwise posterior prediction bands . The figure illustrates adequate fit and predictive coverage for all four subjects.  Over all subjects, the interquartile range (IQR) of predictive coverage rate is  (78.4\%, 88.5\%) and (97.1\% to 100\%) using pointwise prediction bands and simultaneous prediction bands respectively. Similar IQR coverage rates are obtained when considering alpha power dynamics, indicating that the model is reasonably well calibrated \cite{Dawid:1985}. 

A frequentist test for goodness of fit was applied using the framework developed by \cite{Yuan2012}.   Although standardized residual means are consistently close to 0, the p-values for different quantile thresholding scenarios suggest some evidence of lack-of-fit with the residual variance model. A detailed analysis at two levels of the model hierarchy is reported in the supplement \cite{Supplement}. We note that, while these findings are unlikely to influence inference on population summaries, some care would be needed in predictive settings, where a scale mixture extension of the residual model would likely account for the observed overdispersion with respect to Gaussian sampling.

\begin{figure}[h]
	\includegraphics[width = 1.0\linewidth]{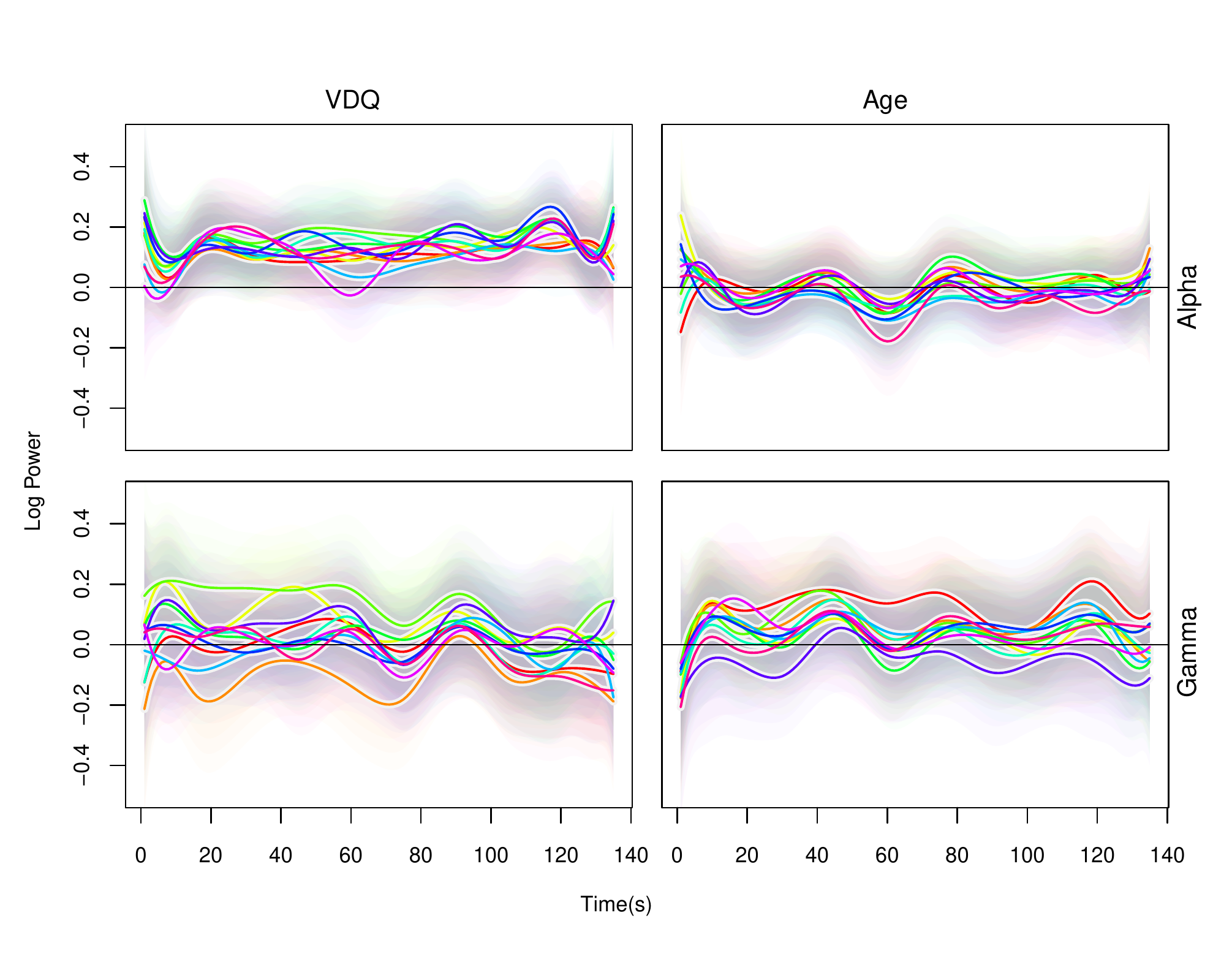}\\
\caption{\small {\bf Varying Coefficients for Age and vDQ.} Varying coefficients associated with age and verbal DQ across all anatomical regions for the Gamma and Alpha frequency bands. Each varying coefficient is accompanied by simultaneous credible bands, shaded to include 0.2, 0.6, and 0.9 of all posterior samples.}
\label{fig:AgeVDQ}
\end{figure}

\subsection{Effects of Age and Verbal-DQ}
\label{sec:age}
Rather than considering a coarse classification of subjects into TD, vASD and mvASD, we examine how alpha and gamma band power trajectories change as a function of age and verbal DQ.  Due to the relatively small-sized sample, the distribution of subjects demographics among three groups is somewhat unbalanced. To be specific, the TD cohort is significantly older ($94.7\pm28.8$, in months) with hider vDQ ($120.6\pm11.4$), the v-ASD cohort are younger ($67.1\pm58$) with medium vDQ ($89.3\pm22.7$) and mv-ASD has a wide age-range ($85.6\pm24.0$), however significantly lower by vDQ ($23.6\pm10.9$). Our analysis is therefore not perfect, but still warranted under the assumption of generalized additive effects.  In particular, the mean structure in \ref{eqn:coeflvl} is represented as the additive combination of time-varying coefficients, including an intercept function, a coefficient function for the main effect of age and a coefficient function for the main effect of vDQ. 

Our results are summarized graphically in Fig.~\ref{fig:AgeVDQ}. We display the varying coefficients for age and vDQ for both the alpha and gamma band power, across all 11 brain regions. For each curve we include probability shading for simultaneous 0.2, 0.6, and 0.9 credible bands. The effects of Age and vDQ on alpha band power is consistent across all brain regions. While age does not seem to be associated with the outcome, higher vDQ levels are consistently and significantly associated with higher alpha band power levels. This finding is consistent with the group-level analysis in Section~\ref{sec:group}, where mvASD children were found to exhibit significantly lower alpha  levels, when compared to TD and vASD children.  On the other hand, the relationship between vDQ and gamma band power is highly heterogeneous across brain regions, with vDQ positively associated with higher gamma band power in temporal regions (LT and RT) and negatively associated with the outcome in the right frontal regions (RF). The effect of Age on gamma power is more consisted across brain regions, with older children exhibiting higher gamma in some regions, and significantly so in LF.

\section{Discussion}
\label{sec:discussion}
We introduced a hierarchical modeling framework for the analysis of region-referenced functional data in the context of functional brain imaging through EEG. Our work hinges on two classical ideas, namely: the constructive definition of Gaussian processes through basis functions, and a flexible representation of regularized dependence within and across anatomical regions through latent factor expansions.  Our proposal is implemented within a high-performance computation \verb'R' package, supporting three prior models, which  include both separable and non-separable covariance structures for region-referenced functional observations. We showed that the proposed approach has satisfactory operating characteristics under extensive numerical experiments, as well as appealing inferential properties, due to straightforward handling of posterior functionals. 

The application of our method relies on projections into the space spanned by basis functions. This feature is engineered into our proposal in order to ensure expert knowledge can be included in the analysis by not limiting the applicability of our method to B-spline projections, but potentially including functional spaces spanned, for example, by wavelets or periodic functions, which may be more appropriate in some applications of multivariate functional data analysis. Specific modeling choices, like the number and placement of spline knots, can be based on straightforward calculation of information criteria \cite{Gelman:2014}.  Similar considerations apply to the choice of prior model, including restrictions on the structure of large covariance oparators. Our implementation depends on choosing the number of latent factors encoding regularized estimation of covariance operators. From our simulation experiments and data analysis, we conclude that regularization through product Gamma priors yields results which are robust to this specific choice. In practice, the number of latent factors included in the analysis can be compared to the number of principal component functions included in a standard FPCA analysis.  A crucial difference in our modeling approach is that we interpret regularized estimation through continuous penalization, rather than discrete truncation to a specific  number of principal components. Therefore, we recommend relative overparametrization,  by selecting a larger number of latent factors in default analyses or by performing formal Bayesian model selection for the number of latent factors \cite{vehtari2017practical}. 

Our case study, considering region-referenced EEG data, shows how complex data structures may be analyzed within the familiar hierarchical modeling framework, coupled with varying coefficient models. Our work focused on inference for the mean structure of region-referenced functions. However, important strides can be made in a formal characterizations of mean and covariance dependence on predictors. This is notably relevant in the field of functional brain imaging, where large levels of subject heterogeneity are often observed. We note that, for the covariance structure, simple group comparisons are indeed possible within the proposed framework, by simply running separate analyses.  Several questions involving regional and functional correlation are, therefore, easily answered from a straightforward examination of posterior summaries. A promising, perhaps more principled, approach would, for example, include a formal representation of covariance heterogeneity through differential subspace structures \cite{Hoff:2016}.

Finally, our work discusses statistical significance only informally, through pairwise comparisons of  simultaneous credible bands within anatomical regions. Within the context of multivariate functional data analysis, more work on the meaning and construction of uncertainty bounds is, however, needed, with formal considerations of multiplicity within structured dependence settings.  

A user-friendly R implementation, with examples, is available in the Supplementary Materials \citep{SupplementCode} and online as an Rcpp package at: \url{https://github.com/Qian-Li/HFM}.

\section*{Acknowledgements}
This work was supported by the grant R01 MH122428-01 (DS, DT) from the National Institute of Mental Health.

\begin{supplement}
\label{suppA}
\stitle{Supplement to ``Region-Referenced Spectral Power Dynamics of EEG Signals: A Hierarchical Modeling Approach''}
\sdescription{
Additional analytical details, extended simulation studies and analyses.
}
\end{supplement}
\begin{supplement}
\label{suppB}
\stitle{Source code for ``Region-Referenced Spectral Power Dynamics of EEG Signals: A Hierarchical Modeling Approach''}
\sdescription{
R package implementation for the methods described in the paper.
}
\end{supplement}

\newpage
\bibliographystyle{imsart-nameyear} 
\bibliography{bib_audio}

\end{document}